\documentclass[a4paper,fleqn]{cas-dc}

\usepackage[authoryear]{natbib}

\usepackage{amsmath,amssymb,amsfonts}
\usepackage{algorithmic}
\usepackage{graphicx}
\usepackage{textcomp}
\usepackage{xcolor}
\usepackage{ragged2e}
\usepackage{balance}
\usepackage{colortbl}
\usepackage{multirow}
\usepackage{balance}
\usepackage[figuresright]{rotating}
\usepackage{rotfloat}
\usepackage{color}

\usepackage{CJKutf8}
\usepackage{framed}

\usepackage{hyperref}
\usepackage{hyperref}

\hypersetup{
    colorlinks=true,
    linkcolor=blue,
    anchorcolor=blue,
    citecolor=blue
}

\def\tsc#1{\csdef{#1}{\textsc{\lowercase{#1}}\xspace}}
\tsc{WGM}
\tsc{QE}
\tsc{EP}
\tsc{PMS}
\tsc{BEC}
\tsc{DE}

\begin{document}
\begin{sloppypar}

\let\WriteBookmarks\relax
\def\floatpagepagefraction{1}
\def\textpagefraction{.001}

\shorttitle{Just-in-Time Multi-Programming-Language Bug Prediction}
\shortauthors{Z Li et al.}
\title [mode = title]{An Exploratory Study on Just-in-Time Multi-Programming-Language Bug Prediction}
    
\author[1]{Zengyang Li}
\ead{zengyangli@ccnu.edu.cn}

\credit{Conceptualization of this study, Methodology, Investigation, Data curation, Writing - Original draft preparation}
\address[1]{School of Computer Science \& Hubei Provincial Key Laboratory of Artificial Intelligence and Smart Learning, \\Central China Normal University, Wuhan, China \\}

\author[1]{Jiabao Ji}
\ead{jjb_coder@mails.ccnu.edu.cn}
\credit{Methodology, Investigation, Data curation, Software, Writing - Original draft preparation}

\author[2]{Peng Liang}
\cormark[1]
\ead{liangp@whu.edu.cn}
\credit{Conceptualization of this study, Methodology, Writing - Original draft preparation}
\address[2]{School of Computer Science, Wuhan University, Wuhan, China}

\author[1]{Ran Mo}
\ead{moran@ccnu.edu.cn}
\credit{Methodology, Writing - Original draft preparation}

\author[3]{Hui Liu}
\ead{hliu@hust.edu.cn}
\credit{Methodology, Writing - Original draft preparation}
\address[3]{School of Artificial Intelligence and Automation, Huazhong University of Science and Technology, Wuhan, China}

\cortext[cor1]{Corresponding author.}

\begin{abstract}
\textbf{}\textbf{Context}: An increasing number of software systems are written in multiple programming languages (PLs), which are called multi-programming-language (MPL) systems. MPL bugs (MPLBs) refers to the bugs whose resolution involves multiple PLs. Despite high complexity of MPLB resolution, there lacks MPLB prediction methods.\\
\textbf{Objective}: This work aims to construct just-in-time (JIT) MPLB prediction models with selected prediction metrics, analyze the significance of the metrics, and then evaluate the performance of cross-project JIT MPLB prediction.\\
\textbf{Method}: We develop JIT MPLB prediction models with the selected metrics using machine learning algorithms and evaluate the models in within-project and cross-project contexts with our constructed dataset based on 18 Apache MPL projects.\\
\textbf{Results}: Random Forest is appropriate for JIT MPLB prediction. Changed LOC of all files, added LOC of all files, and the total number of lines of all files of the project currently are the most crucial metrics in JIT MPLB prediction. The prediction models can be simplified using a few top-ranked metrics. Training on the dataset from multiple projects can yield significantly higher AUC than training on the dataset from a single project for cross-project JIT MPLB prediction.\\
\textbf{Conclusions}: JIT MPLB prediction models can be constructed with the selected set of metrics, which can be reduced to build simplified JIT MPLB prediction models, and cross-project JIT MPLB prediction is feasible.
\end{abstract}


\maketitle

\section{Introduction}
\label{chap:intro}
Software plays a vitally important role in every aspect of today's society, thus software reliability has become increasingly crucial. Detecting or predicting bugs in a timely manner is of great significance for the entire software domain. Traditional defect predictions (i.e., predictions are made late in the
development life cycle) require considerable manpower and incur high costs during software development~\citep{kamei2012large,kamei2016defect}.
Providing practical decision support by predicting bugs in software systems is invaluable for software development teams ~\citep{kamei2016defect}. Software architects and project managers can gain a better understanding of potential future bugs, enabling them to engage in appropriate refactoring activities in advance, thereby reducing future maintenance costs ~\citep{latoza2006maintaining}.

In recent years, there has been a growing interest in just-in-time (JIT) defect prediction because it enables developers to identify changes leading to defects at the time of code commit. Compared to traditional module-level (such as package- or file-level) defect prediction methods, JIT defect prediction is a fine-grained defect prediction technique. It allows developers to examine a smaller number of source code lines to discover potential defects ~\citep{kamei2012large}, which can save a significant amount of effort compared to traditional coarse-grained defect prediction and slow-running defect prediction methods. The capability of JIT defect prediction enables developers to predict defects at the time of code commit, allowing them to quickly inspect and identify potential defects while the details of code changes are still fresh in their minds. As a result, it holds the promise of more efficient detection of potential defects.

Due to the aforementioned benefits of JIT defect prediction, researchers have proposed many methods based on the language-independent change-level features proposed in~\citep{kamei2012large} and conducted research on JIT defect prediction without distinguishing programming languages (PLs)~\citep{liu2017code, zeng2021deep, zhou2022software, cabral2022towards}. Some researchers have also explored JIT defect prediction based on datasets of a specific PL ~\citep{ardimento2022just} or characteristics of a certain PL~\citep{ni2022just}, because the research on single-programming-language (SPL) defect prediction is more targeted and can significantly improve prediction performance.

Modern software systems are predominantly written in multiple PLs ~\citep{kontogiannis2006comprehension, jones1998estimating, mayer2015empirical, li2022exploring, li2023understanding}. Such systems are called multi-programming-language (MPL) systems. Systems developed using multiple PLs can leverage specific language features to achieve certain functionalities, simplify the code, and improve the efficiency of programming ~\citep{abidi2021multi}. With technological advancements, the proportion of projects written in multiple PLs is steadily increasing ~\citep{kontogiannis2006comprehension, li2023understanding2}. Developers utilize the advantages of combining various PLs to cope with the pressure of rapid market updates and iterations. Although software systems offer numerous advantages, they come with increased complexities, including language compatibility issues and more intricate bugs. From the perspective of the number of PLs involved in bug resolution, bugs are divided into two types, i.e., SPL bugs and MPL bugs (MPLBs). The study of MPLBs differs significantly from that of SPL bugs, posing challenges of higher complexity and broader impact~\citep{li2023understanding}. Despite the increasing importance and widespread usage of MPL software systems, to our knowledge, there has been no research on JIT prediction of MPLBs. 

In software development, it is challenging to directly measure whether each commit will introduce bugs. Therefore, we collected relevant metrics of commits and related source code to build our JIT MPLB prediction models. Due to the intricate relationship between commit metrics and bug introduction, linear models are not suitable for JIT MPLB prediction. To fully consider the impact of various metrics on prediction results, we considered four classifiers, including Support Vector Machine (SVM), Logistic Regression
(LR), Decision Trees (DT), and Random Forest (RF), and evaluated their performance respectively. Finally, we chose the RF, which performed the best, to make predictions.~\citep{breiman2001random}. In fact, the bug prediction problem can be viewed as a binary classification task. Our JIT prediction model for MPLBs also takes this into account. 
Through this prediction, developers can quickly assess the likelihood of bug introduction, facilitate better code modifications, and reduce the potential risks associated with these bugs. Our proposed approach can subsequently reduce maintenance costs, making the software development process more efficient and reliable ~\citep{shepperd2014researcher}.

In the early stages of system development, there may be few or no commits, and some systems may have incomplete bug tracking records. This can lead to insufficient training data for the JIT MPLB prediction models. Therefore, we build cross-project JIT MPLB prediction models based on the dataset from other projects. 
This measure is viable because the used metrics exhibit consistency between different projects ~\citep{nagappan2006mining, zimmermann2009cross}, providing an opportunity to investigate cross-project JIT MPLB prediction. We constructed the JIT MPLB prediction models using the Random Forest in machine learning with 23 carefully chosen metrics and 1 commit labels. To train and validate our models, we conducted an empirical study on 18 MPL projects from the Apache open source software (OSS) projects, selected based on project-specific criteria. 

The \textbf{contributions} of this work are: (1) This work is the first attempt to investigate the JIT MPLB prediction problem in MPL projects; (2) We delved into the most significant metrics that impact JIT prediction of MPLBs and constructed streamlined prediction models based on the top-ranked metrics; (3) We evaluated the feasibility of the JIT MPLB prediction models in the cross-project context.

The remainder of this paper is organized as follows. Section \ref{chap:relat} presents the related work. Section \ref{chap:case} describes the experimental design, including metric definition, metric calculation, data preprocessing, model building, model evaluation, and dataset construction. Section \ref{chap:evaluation} describes the research results obtained. Section \ref{chap:discus} discusses research findings. Section  \ref{chap:threats} identifies the threats to validity in our work. Section \ref{conclusions} presents the conclusions and future work.

\section{Related Work}\label{RelatedWork}
\label{chap:relat}

As mentioned in the Introduction section (Section~\ref{chap:intro}), there is currently no research related to JIT MPLB prediction in MPL software systems. Therefore, in this section, we present related work on JIT defect prediction (Section~\ref{JITDP}) and bugs in MPL software systems (Section~\ref{MPLSS}).

\subsection{Just-in-Time Defect Prediction}
\label{JITDP}
The JIT defect prediction methods predict whether a code commit contains defects, identifying risky software changes rather than files or packages. Kamei et al. conducted an empirical evaluation of a JIT quality assurance approach aimed at identifying software changes in real-time that pose a high risk of introducing defects~\citep{kamei2012large}. Recent studies have shown that JIT defect prediction techniques have sufficient predictive accuracy and can be applied in practice~\citep{zhou2022simple,cabral2022towards}.


In the context of SPL defect prediction research, Ni et al. conducted an analysis specific to the JavaScript language. They selected language-independent metrics and five customized JavaScript-specific metrics to perform defect prediction on 20 JavaScript projects. Their findings indicate a significant enhancement in prediction performance when incorporating these relevant language-specific metrics~\citep{ni2022just}. Similarly, Ardimento et al. utilized a dataset compiled from six Java open-source systems and employed a deep temporal convolutional network variant based on hierarchical attention layers to carry out defect prediction. This approach further improved the performance of defect prediction~\citep{ardimento2022just}. Our proposed prediction metrics specific to MPL software projects are also inspired by these relevant studies. By integrating insights from previous research, we aim to develop more effective and tailored metrics for defect prediction in specific PLs.

Existing JIT bug prediction models use the algorithm proposed by Sliwersky, Zimmermann, and Zeller (SZZ)~\citep{sliwerski2005changes} to identify past bug-fixing changes and bug-inducing changes. The SZZ algorithm is commonly employed in defect prediction models. In this paper, we also use the SZZ algorithm (implemented in the PyDriller tool that is adopted in our work) to identify bug-inducing changes~\citep{lin2021impact,yan2020just}.

Cross-project prediction is also a focal point in JIT defect prediction to address issues stemming from insufficient project data. Zeng et al. investigated the performance of DeepJIT, CC2Vec, and other previously studied cross-project validation methods, finding that within-project prediction generally outperforms cross-project prediction~\citep{zeng2021deep}. Kamei et al. proposed that combining data from multiple projects without considering the project background as the training set for cross-project JIT defect prediction can improve defect prediction effectiveness~\citep{kamei2016studying}. We also adopted the approach of merging datasets when performing cross-project prediction and obtained similar conclusions.

Typically, prediction models provide features manually extracted for machine learning classifiers (such as Logistic Regression, Decision  Trees, and Random Forest) to predict whether a code commit is defective~\citep{liu2017code,chen2018multi}. Mo et al. utilized code metrics and history measures as inputs to the Random Forest classification algorithm and achieved good predictive performance~\citep{mo2022exploratory}. Our study also employed similar machine learning methods for JIT defect prediction in experiments.

\subsection{Bugs in MPL Software Systems}
\label{MPLSS}
Various work studied bug proneness in MPL software systems. 
Ray et al. studied the impact of PLs on software quality with a large dataset of 729 OSS projects in 17 PLs collected from GitHub ~\citep{RaPoFiDe2014}. They combined both multiple regression modeling and text analytics to investigate the influence of language characteristics. They found that PL design has a significant but modest effect on software defects. They also found that there is a small but significant correlation between the PL set and software defects. In particular, they found that there are 11 PLs having a relationship with software defects. Berger et al. ~\citep{BeHoMaViVi2019} conducted repeated experiments of the study of Ray et al. ~\citep{RaPoFiDe2014} and reduced the number of defect-related PLs down from 11 to only 4. Li et al. conducted an empirical study on the phenomenon of MPL commits (in which source files written in multiple PLs are changed) in Apache MPL OSS projects to understand the impact of MPL commits on source file bug proneness. They found that in most of the studied projects source files changed in MPL commits are more prone to bugs (by 17.2\% to 3,681.8\%) in terms of defect density than source files that never changed in MPL commits ~\citep{li2021multi, li2022exploring}. Although these studies investigated bug proneness in MPL software systems, they failed to distinguish MPL bugs from SPL bugs and could not provide effective suggestions for MPLB prediction. 

There have been a couple of studies investigating MPLBs in MPL software systems. In 2023, Li et al. used open time, three change complexity measures of pull requests, and two communication complexity measures in bug resolution, to explore the impact of bugs (including MPLBs) on development in three deep learning frameworks ~\citep{li2023understanding2}. 
Li et al. continued their investigation into resolving MPLBs in the MPL software system, as well as the reasons for bugs involving multiple PLs. They conducted a large-scale empirical study on 66,932 bugs collected from 54 MPL projects out of 655 Apache OSS projects. The results showed that: (1) The proportion of MPLBs in the selected projects ranging from 0.17\% to 42.26\%, with an overall project MPLBs proportion of 10.01\%; (2) 95.0\% and 4.5\% of MPLBs involved source files from 2 and 3 PLs, respectively; (3) The change complexity and open time tendency of MPLBs tended to be higher than those of SPL bugs~\citep{li2023understanding}. The both aforementioned studies aim to explore the characteristics of MPLB fixes, which enlightens us to propose MPL metrics to construct the JIT MPLB prediction models.

There have also been a number of studies exploring security MPL vulnerabilities (a specific type of MPL bugs) in MPL software systems. 
Li et al. explored whether MPL code construction holds significant security implications and genuine security consequences through a large-scale study of popular MPL projects and their evolutionary histories on GitHub ~\citep{li2022vulnerability}. They found statistically significant correlations between the vulnerabilities of the MPL code and their PL choices. The correlations were related to the language interface mechanisms rather than individual PLs, indicating a relationship between defects in the MPL code and the language interface mechanisms.
Li et al. proposed a greybox fuzzer that holistically fuzzes a given MPL software system by cross-PL coverage feedback and modeling of the semantic relationships between program inputs and branch predicates across PLs~\citep{li2023polyfuzz}. This fuzzer enabled high code coverage and effective discovery of previously unknown MPL vulnerabilities. Unlike these studies that focus on a certain type of MPL bugs (i.e., MPL vulnerabilities) in MPL software systems, our work investigates MPL bugs in general.




\section{Study Design}
\label{chap:case}
In this section, we describe a set of metrics calculated from commits for JIT MPLB prediction , how to preprocess dataset and how to use these metrics to construct JIT MPLB prediction models. In addition, we present the data collection process to construct the dataset for the evaluation of JIT MPLB prediction.

\subsection{Used Metrics}
In this section, we describe the metrics used for JIT MPLB prediction and explain the calculation of such metrics.
\subsubsection{Metric Definition}
We describe a set of metrics extracted from each commit for JIT MPLB prediction. These metrics are shown in Table \ref{table-metrics}. We classify these metrics into 2 categories. (1) \textbf{Category 1}, which are calculated based on all the modified files in the commit, including all types of files no matter they are supported by the Lizard tool\footnote{https://github.com/terryyin/lizard} or not. Note that all PL files referred to in the rest of this paper are those supported by Lizard. (2) \textbf{Category 2}, which are calculated based on the source files written in the 18 general-purpose PLs supported by the Lizard tool, as shown in Table \ref{Lizard-support-language}. In the following, we explain the two categories of metrics in detail.

\textbf{Category 1:} As shown in Table \ref{table-metrics}, metrics C1-C7 were collected from the following four perspectives. (1) The complexity of the current commit ~\citep{moser2008comparative, nagappan2005use}: number of modified lines of code and source files in the current commit (C1-C4). (2) The project's complexity at the time of the current commit ~\citep{menzies2002metrics}: accumulated number of source files and lines of code modified in previous commits before the current commit (C5-C6). (3) The familiarity of the author with the source files being modified in the current commit: determining whether the same author made consecutive modifications to source files (C7).


\textbf{Category 2: }As shown in Table \ref{table-metrics}, metrics C8-C23 were collected from four perspectives: (1) From the selected MPL files (all PL files modified in one single commit), we calculated the number of lines added, deleted, and files modified in the commit, as well as whether it was a modification in the primary PL (C8-C11). This intuitive metric assesses the MPL complexity of this commit. (2) The complexity of the commit within the modified source files: the unit size property, the unit complexity property, and the unit interface property (C12-C14) obtained from detailed analysis of the source files in the commit. (3) Further analysis was conducted on the specific source code in MPL files, considering the maximum and average values of metrics such as the previous method count, the current modified method count, cyclomatic complexity, and the token count (C15-C22). These measurements were taken from both an overall and a peak perspective to evaluate the modification complexity at the source code level. (4) Calculation of the entropy of MPL files (C23) to measure the disorderliness of MPL files~\citep{hassan2009predicting}. Note that if a commit involves source files in only one single PL, these metrics can be calculated based on the source files in this PL.

\begin{table}[]
\caption{Metrics for each commit}
\centering
\begin{tabular}{p{0.06\columnwidth} p{0.85\columnwidth} }
\toprule
\textbf{Index}& \textbf{Metric description}  \\ \midrule
C1& Deleted lines of code (LOC) of all files\\
C2& Added LOC of all files\\
C3& Changed LOC of all files\\
C4& Number of all modified files\\
C5& The total number of lines of all files of the project currently \\
C6& Total number of all files of the project currently \\
C7& The proportion of files modified consecutively by the same author to the total number of files modified in this commit\\
C8& Deleted LOC of MPL files\\
C9& Added LOC of MPL files\\
C10& Number of modified MPL files\\
C11& Whether to modify the source files of the main PL\\
C12& Delta Maintainability Model (DMM) metric value for the unit (e.g., method) size property~\citep{di2019delta}\\ 
C13& DMM metric value for the unit complexity property \\
C14& DMM metric value for the unit interfacing property \\
C15& Average cyclomatic complexity of the modified MPL files~\citep{Shepperd1988}\\
C16& Average number of methods of the modified MPL files in this commit before this modification\\
C17& Average number of methods modified in MPL files\\
C18& Average token count in the modified MPL files (In PyDriller, a ``token'' refers to a basic unit of source code, such as a keyword, identifier, operator, or punctuation symbol)\\
C19& Maximum cyclomatic complexity of the modified MPL files\\
C20& Maximum number of methods of the modified MPL files in this commit before this modification\\
C21& Maximum number of methods modified in the MPL files\\
C22& Maximum token count in the modified MPL files\\
C23& Entropy of the modified MPL files \\
\bottomrule
\end{tabular}
\label{table-metrics}
\end{table}

\subsubsection{Metric Calculation}\label{MetricCalculation} 
Most metrics are straightforward, thus, we only explain the calculation of several relatively complex metrics.\\
\textbf{C6:} Obtain the last author who modified each file, calculate the proportion of files modified by the same author in two consecutive modifications out of the total number of modified files.\\
\textbf{C12-C14:} The DMM metric is the proportion of low-risk changes in the commit. The resulting values range from 0.0 (all changes are risky) to 1.0 (all changes are low risk). It rewards making methods better and penalizes making things worse. The DMM implementation expands on the PyDriller metrics by incorporating three commit-level metrics that assess risk factors related to size, complexity, and interfacing\footnote{https://pydriller.readthedocs.io/en/latest/deltamaintainability.html}.\\
\textbf{C8-C23: }
Most metrics are calculated based on the output of PyDriller that depends on Lizard; thus, in this study, we only use the source files written in the PLs supported by Lizard.   \\
\textbf{C23:} Extract MPL file extensions in this commit, calculate the number of files in each PL, and use formula $H(m) = -\sum_{i=1}^mp_ilog_2 p_i$ to determine the file entropy of this commit, where $m$ represents the language type, and $p_i$ represents the proportion of individual language file count to the total count of all language files~\citep{hassan2009predicting}.\\

\begin{table}[]
\caption{Programming languages supported by Lizard}
\centering
\begin{tabular}{cccc}
\toprule
\textbf{Index} & \textbf{Language} & \textbf{Index} & \textbf{Language} \\ \midrule
L1             & C             & L10            &   Lua    \\
L2             & C++           & L11            &   Objective-C          \\
L3             & C\#           & L12            &   PHP          \\
L4             & Erlang        & L13            &   Python          \\
L5             & Fortran       & L14            &   Ruby            \\
L6             & Go            & L15            &   Rust       \\
L7             & Java          &L16             &   Scala      \\
L8             & JavaScript    & L17            &   Swift             \\
L9             & Kotlin        &L18             &   TypeScript    
\\\bottomrule
\end{tabular}
\label{Lizard-support-language}
\end{table}

\subsection{Commit Labeling}\label{Labeling} 
To perform JIT MPLB prediction, for each commit we need to label whether it introduces an MPLB. This commit label is described as whether the commit introduces one or more bugs and is fixed by MPL files in a commit, and we name it P1. If a commit (\textit{commit\_a}) introduces a bug and this bug is fixed in an MPL commit (\textit{commit\_b}), then we consider that \textit{commit\_a} introduces an MPLB.

This commit label is tagged based on the source files modified in the commit, related bugs, and their bug-fixing commits. The tagging process of commit label P1 for a commit is presented below:

\noindent \textbf{Step1:} By matching the messages of commits with the key values in Jira's issue table using regular expressions, specifically by matching with bug ticket IDs, we identified the commits related to bug fixes. Utilizing the \textsc{get\_commits\_last\_modified\_lines} function from the PyDriller library, we obtained commits that modified files from the original commit. We define this commit as the one that causes the bug. Similar methods have been employed in existing work to identify changes made for bug fixes~\citep{kamei2012large, kim2008classifying, mo2022exploratory}.\\
\noindent \textbf{Step2:} If the bug is fixed in a commit in which the modified source files written in two or more general-purpose PLs supported by Lizard, then P1 is true.


\subsection{Data Preprocessing}
\textbf{Data Normalization: }In our dataset, there are metrics such as C12-C14, which range from 0 to 1, and metrics like C5, which can be hundreds of thousands. The significant differences in data scales among these metrics may lead to gradient vanishing or exploding issues, as well as increase the complexity of the model. To address these problems, we employ Min-Max normalization techniques to unify the data scales. When $X$ represents a column of data, the formula for Min-Max normalization is as follows:
\begin{equation} 
X_{\text{norm}} = \frac{X - \text{min}(X)}{\text{max}(X) - \text{min}(X)}  
\end{equation}  

In cross-project prediction experiments, to ensure data consistency, the normalization of the test set data is performed using the normalization parameters obtained from the training set.

\textbf{Dealing with the Imbalanced Dataset: }Our data is imbalanced, meaning that commits introducing MPLBs only account for a small portion of all commits. If handled improperly, this imbalanced data can lead to a decrease in the performance of predictive models. To address this data imbalance issue, we adopt an undersampling approach to process the data. Specifically, we randomly reduce the number of samples from the majority class (commits that do not cause MPLB) so that the majority class samples are reduced to the same level as the minority class (commits that cause MPLB).

\subsection{Building and Evaluating the Prediction Models}\label{BuildEvaluationModel}
After our investigation, we found that there were no relevant methods for predicting MPLBs in previous studies on defect prediction. Due to the different prediction objects of those studies from our work, we cannot directly use and compare the defect prediction methods in the previous studies. Under such conditions, we have decided to explore the prediction of MPLB using traditional classifiers in machine learning. To establish the best-performing model for JIT MPLB prediction, we further investigated the use of four additional algorithms, i.e., Support Vector Machine, Logistic Regression, Decision Trees, and Random Forest, to build the prediction models. These algorithms are widely employed in bug prediction~\citep{kamei2012large,mo2022exploratory}.

\textbf{Support Vector Machine (SVM): } SVM is a supervised learning algorithm that has gained significant popularity in various machine learning applications due to its high accuracy and ability to handle high-dimensional data. The core idea of SVM is to find a hyperplane in a multi-dimensional space that can distinguish data points of different classes with the maximum margin~\citep{hearst1998support}.

\textbf{Logistic Regression (LR): }LR is a classification method that uses a sigmoid function to convert linear regression outputs into probabilities for binary classification. It's simple, efficient, and interpretable, but assumes a linear relationship between features and target variables~\citep{hosmer2013applied}.

\textbf{Decision Trees (DT): }DT classify data by breaking it down into increasingly specific branches, ending in decisions based on the features of the data. They are simple yet powerful~\citep{kotsiantis2013decision}.

\textbf{Random Forest (RF): }RF is a supervised machine learning algorithm, and it combines the concepts of decision trees and ensemble learning. Random Forest is an ensemble model composed of multiple decision trees. Each decision tree serves as an independent classifier, and the final classification or regression prediction is made through voting or averaging the individual predictions~\citep{quinlan1986induction}. It offers higher accuracy than a single decision tree and helps reduce overfitting. 

The classification threshold for all four classifiers in the experiment is set to 0.5, which means that when the model predicts the probability of an MPLB for a commit to be greater than 0.5, this commit will be classified as a commit that leads to the MPLB. Otherwise, it will be classified as a commit that does not lead to the MPLB.

\textbf{Model Training and Validation: }In the training and validation of a single project, we employed a ten-fold cross-validation technique to ensure the stability of the prediction model and select an appropriate prediction model. Cross-validation is a widely used method in the field of machine learning~\citep{stonecross}, which estimates the performance of prediction models in real-world applications by dividing the original dataset into training and testing sets. In our study, we chose the ten-fold cross-validation method. In this technique, the original dataset is first randomly divided into ten equally sized subsets. Subsequently, the process is iterated ten times. During each iteration, nine subsets are utilized as training data, and the remaining one subset is used as validation data. This means that each subset is used as validation data once and as training data nine times. At the end of each iteration, we obtain an estimate of the model's performance. The advantage of this cross-validation technique lies in its effective utilization of all available data. In multiple iterations, each data point is used for both training and testing, providing a more comprehensive evaluation of the model's performance. 

In cross-project experiments, we used the data from other individual projects or the combined data from all other projects as the training set, and the data from the current project as the test set to train and validate the model. Since our aim is to assess the model's ability to predict MPLBs across projects, cross-validation is not applicable in this scenario.

\textbf{Model Evaluation:} Our models are binary classification models, hence we use precision, recall, and F1-score to measure the performance of the bug prediction models~\citep{sokolova2009systematic}. Precision measures the accuracy of the model's predictions within each category, recall tells how much proportion of actual samples in each category the model successfully identified, and finally, the F1-score presents a comprehensive view by combining precision and recall. True Positive, False Positive, and False Negative are abbreviated as TP, FP, and FN, respectively. The formulas for precision, recall, and F1-score are given as follows:

\begin{equation}
Precision = \frac{{TP}}{{TP}+{FP}}
\end{equation}
\begin{equation}
Recall = \frac{{TP}}{{TP}+{FN}}
\end{equation}
\begin{equation}
F1-Score = \frac{2\times Precision\times Recall}{Precision+Recall}
\end{equation}

The aforementioned performance metrics, such as precision and recall, may vary when adjusting the threshold. To obtain an overall concept of performance across different thresholds, we utilize the area under the curve (AUC) of the receiver operating characteristics (ROC)~\citep{lobo2008auc}. The AUC ranges from 0 to 1, where a value closer to 1 indicates better prediction performance. Any predictor with an AUC value greater than 0.5 is more effective than a random predictor. The advantage of the AUC of ROC is its robustness to imbalanced positive and negative samples, as the ROC curve is plotted by traversing all possible classification thresholds, taking into account all possible classification results. When dealing with imbalanced datasets, selecting an appropriate classification threshold can be challenging, but AUC provides us with an evaluation standard independent of the threshold.

The results presented in this article are the mean values obtained from cross-validation. Through these performance metrics, we can thoroughly evaluate the performance of the classification model, providing us with a better understanding of the model's behavior. These metrics are commonly used in research papers to assess the performance of predictions as well~\citep{zhou2022software, mo2022exploratory}.

\subsection{Dataset Construction}\label{Dataset}
In this section, we first present the criteria for selecting appropriate MPL projects serving as subjects to evaluate our JIT MPLB prediction approach, and then describe the procedure to collect the needed data.

\subsubsection{Project Selection}
\label{projectselection}
In this study, we selected MPL software projects supported by the Apache Software Foundation. We adopted Apache projects because the links between issues and corresponding commits tend to be well maintained in the commit messages of these projects. To determine the MPL projects included in our study, we applied the following selection criteria:

\begin{itemize}
\setlength{\itemsep}{0pt}\setlength{\parskip}{0pt}
  \item [1)] 
  In terms of PLs, we chose to analyze MPL projects. Each project uses at least two PLs out of the 18 general-purpose PLs supported by the Lizard tool, which was used to calculate C8-C23. The PLs supported by Lizard are shown in Table \ref{Lizard-support-language}. The code written in each of the two most prominent PLs should constitute more than 5\% of all the source code of the project, and the proportion of code in the main PL should be less than 90\%.      
  \item [2)]
 In terms of project evolution, we chose to include projects with at least 1000 commits.  
  \item [3)]
  In terms of project activity, we selected projects that had commits in the past year. 
  \item [4)]
    In terms of bug collection, we chose projects that manage issues by Jira, which can use Jira REST API to conveniently export issue data.
\item [5)]
The percentage of bug-inducing commits should be 10\% or more of all commits. If the percentage of bug-inducing commits is too small, there will be a severe class imbalance problem, and it will be impossible to train a reasonable bug prediction model~\citep{batista2004study}. Note that after excluding such projects with low percentages of bug-inducing commits, we still performed undersampling on the imbalanced dataset of the remaining MPL projects.
\end{itemize}


\subsubsection{Data Collection Procedure} \label{Collection-steps}
\begin{figure*}[]
  \centering
  \includegraphics[width=0.7\textwidth]{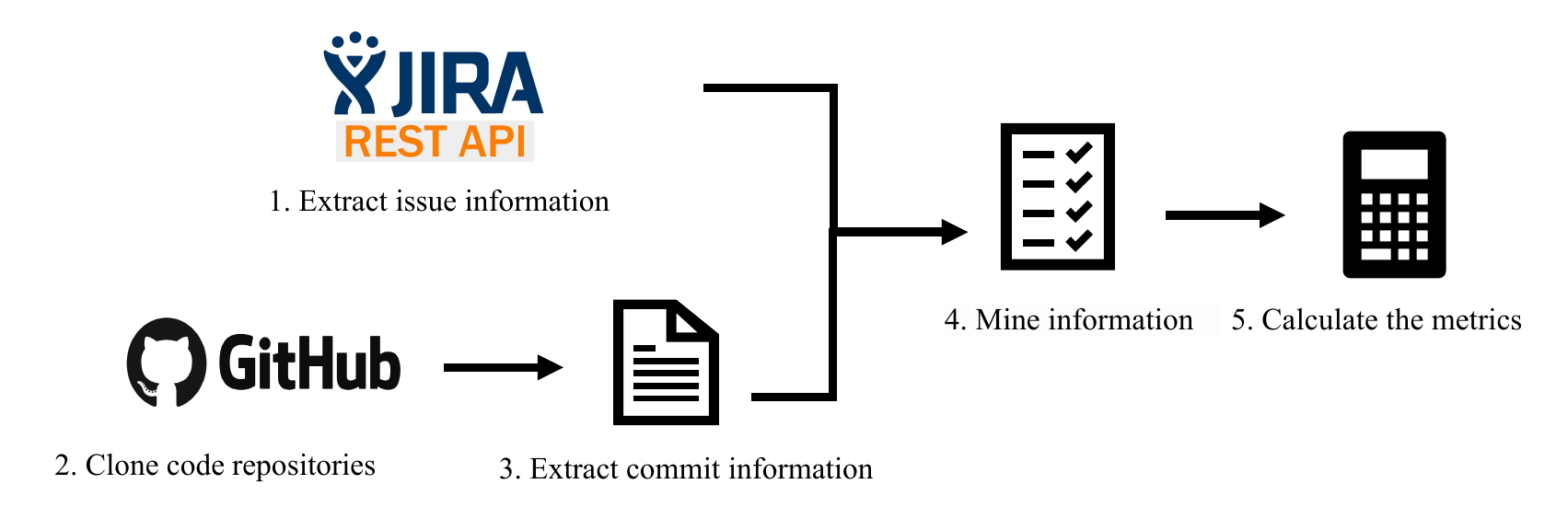}\\
  \caption{Procedure of data collection}
  \label{figure-data-process}
\end{figure*}

The data were collected around July 30th, 2022.
As shown in Figure \ref{figure-data-process}, the data collection procedure for each selected project consists of 5 steps.\\
\noindent \textbf{Step 1}: Extracted issue information (e.g., issue type and issue key for each issue) from  Jira\footnote{https://www.atlassian.com/software/jira} issue tracking system through the Jira REST API\footnote{https://developer.atlassian.com}.\\
\noindent \textbf{Step 2}: Cloned the repository of the  project from the master branch on GitHub to our local environment. This significantly speeds up the data analysis process. We utilized the master branch as it contains the primary development version of the project.\\
\noindent \textbf{Step 3}: Extracted commit information from the GitHub repository of the project.\\
\noindent \textbf{Step 4}: Used PyDriller to mine information related to the used metrics in Table \ref{table-metrics} and the commit labels from the cloned repository. PyDriller was chosen for mining information from GitHub repositories due to its significant advantages with relatively low complexity for mining Git repositories~\citep{spadini2018pydriller}.

\noindent \textbf{Step 5}: Calculated the metrics described in Table \ref{table-metrics} and tagged the commit labels for each commit through a calculation tool that we developed.

\section{Evaluation}
\label{chap:evaluation}

\subsection{Research Questions}
In this article, we conducted exploratory research by investigating the following research questions (RQs):

\textbf{RQ1: What is the performance of JIT MPLB prediction based on the metrics?}

\textbf{Rationale:} First, we used four commonly used defect prediction classifiers, i.e., SVM, Logistic Regression, Decision Trees, and Random Forest, to predict the dataset and obtained the best prediction model. Then, we further explored the best prediction model.
 Despite the high popularity and importance of MPL software systems, research on bug prediction in MPL projects is lacking due to the complexity of MPL projects~\citep{mayer2015empirical}. JIT MPLB prediction provides timely insight and information, enabling the identification of MPLBs in the early stages of code commit. It offers developers relevant suggestions, aiding them in making informed decisions promptly, reducing time cost, and minimizing the workload for subsequent maintenance and fixes. Therefore, a practical JIT MPLB prediction model is crucial.

\textbf{RQ2: Which metrics are the most important on JIT MPLB prediction?}

\textbf{Rationale:} With this RQ, we explored which metrics significantly influence JIT MPLB prediction, i.e., the metrics that contribute the most to JIT MPLB prediction. To this end, we extensively gathered relevant variables (i.e., metrics) that could have a significant impact on bug prediction. Testing the Random Forest prediction model with these variables, we determined the combination of metrics that had the most significant impact on the model's prediction performance based on the results. 
The metrics we collected can also provide valuable insights for relevant practitioners, enabling them to consider and enrich their research on the metrics. This approach helps in constructing the best possible model from a methodological perspective.

\textbf{RQ3: When the metrics used are reduced, how does the performance of the JIT MPLB prediction model change?}

\textbf{Rationale:} After studying the significance of the metrics in the previous RQ, we further investigated whether these metrics could be reduced. This reduction would enable researchers and practitioners to make similar predictions without the need for such an extensive set of metrics. Up to this point, we have defined 23 metrics and 1 commit labels, many of which involve calculations related to the core complexity of the code. There is no doubt that when more metrics are taken into consideration in a prediction model, additional effort and cost must be devoted to data acquisition and processing. Although including more information may improve prediction performance to some extent, it inevitably requires a greater investment of resources~\citep{he2015empirical}. Therefore, a simplified model with fewer metrics but similar performance could allow research teams to adopt our method more easily and make predictions more efficiently.

\textbf{RQ4: What is the performance of cross-project JIT MPLB prediction?}

\textbf{Rationale:} We explore this RQ because some projects are poorly labeled with information, which may lead to missing data. Consequently, there might be insufficient bug annotation information or an inability to obtain bug annotations at all. This situation hampers the construction of effective models for the prediction of JIT MPLB. Therefore, we conduct cross-project predictions to address the aforementioned issues.

\subsection{Selected Projects and Constructed Dataset}

\begin{table*}
\caption{Demographic information of the selected projects}
\scalebox{0.8}{
\centering
\begin{tabular}{crrrrrrrrrrrrr}
\toprule
\textbf{Project} & \textbf{\#Star} & \textbf{\#Commit} & \textbf{\#Ctbtr} & \textbf{Age(y)} & \textbf{\#BIC} & \textbf{\%BIC} & \textbf{LOC} & \textbf{\#File} & \textbf{\#PL} & \textbf{Coverage} & \textbf{Language\_1} & \textbf{Langauge\_2} \\ \midrule
Airavata         & 89               & 10,105             & 46                     & 10                 & 1,349           & 13.3         & 21,678,436     & 12,974       &6 &96.7\%   & Java: 75.4\%         & C++: 16.8\%          \\
Ambari           & 1,985            & 24,864             & 146                    & 12                 & 15,352          & 61.7         & 34,177,067     & 32,168       &7 &84.4\%   & Java: 44.6\%         & JavaScript: 24.4\%   \\
Atlas            & 1,641            & 3,776              & 122                    & 6                  & 1,840           & 48.7         & 1,914,028      & 4,370        &3 &94.0\%   & Java: 59.6\%         & JavaScript: 31.6\%   \\
Avro             & 2,572            & 3,889              & 295                    & 15                 & 790             & 20.3         & 1,135,209      & 3,101        &9 &96.5\%   & Java: 42.3\%         & C\#: 15.5\%          \\
CloudStack       & 1,506            & 36,137             & 214                    & 11                 & 7,048           & 19.5         & 379,881        & 322          &5 &93.3\%   & Java: 67.1\%         & Python: 20.9\%       \\
Hawq             & 693              & 1,772              & 65                     & 8                  & 437             & 24.7         & 10,568,907     & 7,898        &5 &91.4\%   & C: 70.3\%            & Python: 9.0\%        \\
Ignite           & 4,527            & 28,659             & 249                    & 9                  & 6,561           & 22.9         & 44,847,804     & 25,570       &7 &99.2\%   & Java: 79.1\%         & C\#: 12.3\%          \\
Impala           & 1,002            & 10,955             & 160                    & 8                  & 4,643           & 42.4         & 17,266,907     & 7,449        &5 &96.6\%   & C++: 51.0\%          & Java: 26.4\%         \\
Kafka            & 25,773           & 11,632             & 349                    & 12                 & 4,504           & 38.7         & 2,924,576      & 6,346        &3 &99.6\%   & Java: 77.1\%         & Scala: 20.2\%        \\
Kudu             & 1,746            & 10,083             & 118                    & 8                  & 2,234           & 22.2         & 1,756,813      & 2,836        &6 &96.6\%   & C++: 79.2\%          & Java: 10.6\%         \\
Kylin            & 3,519            & 8,793              & 192                    & 9                  & 2,151           & 24.5         & 17,456,602     & 14,822       &6 &83.8\%   & Java: 59.0\%         & JavaScript: 13.8\%   \\
Mahout           & 2,092            & 4,555              & 42                     & 10                 & 1,363           & 29.9         & 3,106,936      & 5,942        &4 &94.7\%   & Java: 80.7\%         & Scala: 14.0\%        \\
Nifi             & 3,976            & 8,800              & 318                    & 9                  & 2,731           & 31.0         & 4,789,176      & 14,240       &6 &96.1\%   & Java: 89.1\%         & JavaScript: 6.7\%    \\
Ranger           & 781              & 4,316              & 119                    & 9                  & 1,764           & 40.9         & 2,088,231      & 4,258        &4 &89.8\%   & Java: 65.6\%         & JavaScript: 19.4\%   \\
Samza            & 782              & 2,585              & 131                    & 9                  & 873             & 33.8         & 1,142,603      & 2,936        &4 &99.1\%   & Java: 86.7\%         & Scala: 11.5\%        \\
Storm            & 6,477            & 10,667             & 278                    & 10                 & 1,289           & 12.1         & 6,091,882      & 4,522        &5 &93.3\%   & Java: 83.6\%         & Python: 7.5\%        \\
Thrift           & 9,896            & 6,830              & 322                    & 14                 & 2,026           & 29.7         & 1,506,153      & 3,794        &15&77.6\%   & C++: 32.9\%          & Java: 9.4\%          \\
ZooKeeper        & 11,528           & 2,510              & 191                    & 15                 & 873             & 34.8         & 883,504        & 2,154        &5 &92.4\%   & Java: 74.2\%         & C++: 7.4\%           \\\bottomrule 
\end{tabular}
}
\label{table-project-info}
\end{table*}

As shown in Table \ref{table-project-info}, following the project selection criteria, we selected 18 MPL projects from all the 655 Apache projects. These projects utilize multiple PLs and have diverse functionalities. Their functions are briefly described in Table \ref{table-project-description} in the Appendix.

Table \ref{table-project-info} shows the demographic information of the selected projects. \#Star is the number of stars, ranging from 89 to 25,773 for the projects. \#Commit is the number of commits, with 8 projects having over 10,000 commits and 3 projects having less than 3,000 commits. \#Ctbtr is the number of contributors, ranging from 42 to 349 for the projects. Age is the duration of the project, with all the projects over 6 years old. \#BIC is the number of bug-inducing commits in each project, ranging from 437 to 15,352 for the projects. \%BIC is the percentage of bug-inducing commits over all commits, with the smallest of 12.1\% for project \textit{Storm} and the greatest of 61.7 for project \textit{Ambari}. LOC is the total number of lines of code in the project, and LOC for the projects ranges from around 380K to 44,848K, indicating all the projects are large-scale. \#File is the total number of source files in the project, falling in the range of [322, 32,168]; \#PL is the number of general-purpose PLs supported by Lizard, with the least of 3 for \textit{Atlas} and the greatest of 15 for \textit{Thrift}; Coverage is the percentage of source code written in all the general-purpose PLs that Lizard supports, with all projects over 77.6\% and 14 projects over 90.0\%; Language\_1 and Language\_2 columns show the top two most used PLs for each project, respectively, and it can be observed that Java and C++ as the main PLs account for the majority of the projects.

Our study collected 23 prediction metrics, providing a convenient resource for researchers exploring JIT MPLB prediction. We labeled bug-inducing commits for 36 projects that were left by the first four screening rules in Section \ref{projectselection}. Furthermore, we collected the prediction metrics, as well as commit labels from related commit records, bugs, and source files, for the 18 finally selected Apache MPL OSS projects. We have shared this dataset and the relevant code~\citep{dataset} for replication purposes.



\subsection{Study Results}

\subsubsection{Performance of JIT MPLB prediction (RQ1)}

\begin{table}[]
\caption{AUC values for the JIT MPLB prediction results using various algorithms (RQ1)}
\centering
\begin{tabular}{ccccc}
\toprule
         \textbf{Project} & \textbf{DT (\%)}& \textbf{LR (\%)} & \textbf{SVM (\%)} & \textbf{RF (\%)} \\\midrule
        Airavata & 84.78 & 84.64 & 84.99  & \textbf{89.84}\\ 
        Ambari & 64.97 & 66.74 & 63.96  & \textbf{73.36}\\ 
        Atlas & 72.73 & 73.66 & 73.76  & \textbf{79.97}\\ 
        Avro & 65.50 & \textbf{67.17} & 66.58  & 66.58\\ 
        CloudStack & 67.59 & 64.72 & 64.79  & \textbf{75.87}\\ 
        Hawq & 65.92 & 65.67 & 62.33  & \textbf{70.25}\\ 
        Ignite & 64.70 & 67.44 & 65.17  & \textbf{73.06}\\ 
        Impala & 62.70 & 69.37 & 69.35  & \textbf{72.84}\\ 
        Kafka & 66.30 & 70.12 & 69.25  & \textbf{75.03}\\ 
        Kudu & 66.37 & 63.48 & 64.50  & \textbf{74.80}\\ 
        Kylin & 64.39 & 63.54 & 64.45  & \textbf{76.42}\\ 
        Mahout & 69.49 & 69.16 & 70.49  & \textbf{73.92}\\ 
        Nifi & 76.44 & 80.40 & 79.74  & \textbf{82.48}\\ 
        Ranger & 73.91 & 74.54 & 74.75  & \textbf{77.05}\\ 
        Samza & 68.02 & 73.87 & 74.63  & \textbf{75.99}\\ 
        Storm & 64.86 & 64.07 & 63.44  & \textbf{75.94}\\ 
        Thrift & 66.66 & 68.90 & 68.12  & \textbf{72.79}\\ 
        ZooKeeper & 72.47 & 73.51 & 69.97   & \textbf{84.21} \\         Average &68.77&70.06&69.46 &\textbf{76.13}\\\bottomrule
\end{tabular}
\label{table-mult-methods-auc}
\end{table}

\begin{table}[]
\caption{Evaluation results of JIT MPLB prediction models using Random Forest (RQ1)}
\centering
\begin{tabular}{cccc}
\toprule
\textbf{Project} & \textbf{Precision\%} & \textbf{Recall\%} & \textbf{F1-Score\%} \\\midrule
Airavata & 90.88 & 87.84 & 89.09 \\ 
        Ambari & 74.24 & 71.46 & 72.81 \\ 
        Atlas & 80.18 & 79.63 & 79.72 \\ 
        Avro & 66.67 & 71.50 & 66.27 \\ 
        CloudStack & 76.50 & 74.14 & 75.21 \\ 
        Hawq & 72.17 & 68.83 & 67.12 \\ 
        Ignite & 74.01 & 71.12 & 72.45 \\ 
        Impala & 74.97 & 68.66 & 71.64 \\ 
        Kafka & 75.57 & 74.01 & 74.74 \\ 
        Kudu & 75.66 & 74.94 & 74.89 \\ 
        Kylin & 77.59 & 73.39 & 74.70 \\ 
        Mahout & 74.27 & 73.71 & 73.73 \\ 
        Nifi & 81.77 & 81.23 & 81.26 \\ 
        Ranger & 77.42 & 76.83 & 76.93 \\ 
        Samza & 76.41 & 74.18 & 75.04 \\ 
        Storm & 74.17 & 76.17 & 72.17 \\ 
        Thrift & 72.38 & 73.02 & 72.26 \\ 
        ZooKeeper & 83.57 & 84.82 & 83.82 \\    
        Average & 76.58 & 75.30 & 75.21
        \\\bottomrule   
\end{tabular} 
\label{table-prediction-results}
\end{table}

To address RQ1, we constructed a bug prediction model based on the 23 metrics and 1 commit label (as shown in Table \ref{table-metrics}) using SVM, Logistic Regression, Decision  Trees and Random Forest. We utilized 90\% of the data as the training set and 10\% as the test set to build JIT MPLB prediction models. We measured the model's prediction performance AUC.

As shown in Table \ref{table-mult-methods-auc}, we used bold font to highlight the best-performing data for each project and can observe that the Random Forest classifier achieves the highest average AUC value of 76.13\% compared to the other three classifiers. Specifically, Random Forest performs the best for 17 out of the 18 projects and the second best for the rest one project. This finding indicates that Random Forest is the most effective classifier for prediction. Therefore, for subsequent experiments in this study, we will use Random Forest as the classifier.

As shown in Table \ref{table-mult-methods-auc} and Table \ref{table-prediction-results}, we observed the following: (1) The projects achieved an average precision of 76.58\% (ranging from 66.67\% to 90.88\%) with only one project (i.e., \textit{Avro}) below 70.00\%, showing relatively stable precision. (2) In terms of recall, the average recall of the projects is 75.30\%, with project \textit{Airavata} reaching 87.84\% and two projects below 70.00\%. (3) Regarding F1-score, the majority of the projects were around 75\%, with project \textit{Airavata} reaching an impressive recall rate of 89.09\%. (4) The average AUC of all projects is 76.13\%, with project \textit{Airavata} reaching 89.84\%. These results suggest that the Random Forest prediction models performed well in all the four performance metrics.

\begin{center}
\fcolorbox{black}{gray!10}{\parbox{0.97\linewidth}{\textbf{Answer to RQ1:} Using the selected prediction metrics, the Random Forest in machine learning enables relatively accurate and stable JIT prediction of MPLBs. By using the Random Forest, we have obtained the average values of precision, recall, F1-score, and AUC for 18 projects, which are 76.58\%, 75.20\%, 75.21\%, and 76.13\%, respectively.}}
\end{center}

\subsubsection{Metric importance on JIT MPLB prediction (RQ2)}

\begin{figure*}[]
  \centering
  \includegraphics[width=0.96\textwidth]{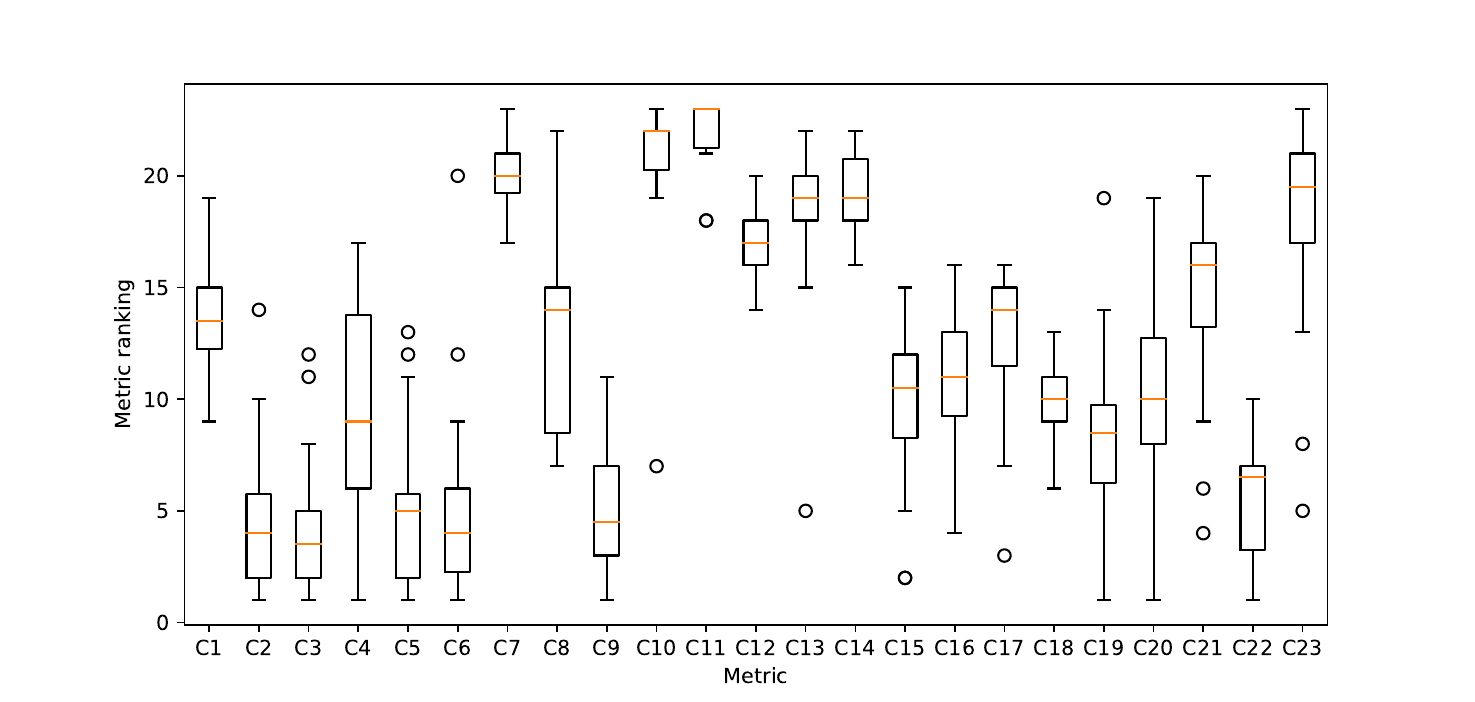}\\
  \caption{Importance level distribution of the
  metrics (RQ2)}
  \label{figure-index-importance}
\end{figure*}

\begin{table}[]
\caption{Ranking of metrics (RQ2)}
\centering
\begin{tabular}{cccccc}
\toprule
\textbf{Metric} & \textbf{Rank} & \textbf{Metric} & \textbf{Rank} & \textbf{Metric} & \textbf{Rank} \\\midrule
C3            & 4.22          & C15             & 9.50          & C12            & 17.10         \\
C2            & 4.78          & C18            & 9.61         & C23           & 17.89         \\
C5             & 4.78          & C20            & 10.06         & C13           & 18.22         \\
C9             & 4.89          & C16            & 10.78         & C14           & 19.28         \\
C6             & 5.39          & C8            & 12.83         & C7            & 20.17         \\
C22             & 5.83          & C17            & 12.89         & C10           & 20.61         \\
C19             & 8.06          & C1             & 13.33         & C11           & 22.06         \\
C4            & 9.39          & C21            & 14.33         &               &               \\
\bottomrule    
\end{tabular}
\label{table-prediction-metrics-rank}
\end{table}


To resolve RQ2, we calculated the Gini importance~\citep{nembrini2018revival} of each feature in tree-based models to measure the significance of various metrics. During the training of each decision tree, random forest tracks the reduction in Gini impurity for each feature. Gini impurity is a metric used to measure the impurity of a dataset. In the context of decision trees, the Gini impurity represents the probability that a randomly selected sample is incorrectly classified. When a feature is used to split a node, it gains a Gini importance score based on the reduction in Gini impurity. Specifically, the Gini importance of each feature is calculated by adding the reduction in Gini impurity for that feature across all trees in the Random Forest. Typically, after obtaining Gini importance scores, the scores are normalized to ensure that the sum of importance scores for all features is equal to 1. This normalization allows a direct comparison of the relative importance of different features.

For each project's prediction, we first computed the Gini impurity for the provided metrics. Subsequently, we ranked the metrics from 1 to $N$ according to their importance scores, where $N$ is the total number of metrics (23 in this case). The metric with the highest Gini importance was ranked 1, while the metric with the lowest Gini importance was ranked as 23. A higher ranking indicates that the metric provides more useful information. We trained the Random Forest model using the entire dataset for each project and calculated the results separately for 18 projects, obtaining Fig. \ref{figure-index-importance} which shows the distribution of rankings for the metrics.

From Fig. \ref{figure-index-importance}, we found that (1) the rankings of each metric vary across the studied projects, and (2) the ranking distributions of various metrics differ among the projects. For instance, C1 (Total lines deleted) of the projects was ranked between 9th and 19th among all metrics, and C9 (Added LOC of MPL files) of the projects was ranked between 1st and 11th among all metrics.

To further explore the metrics intuitively, we calculated the average importance values for each metric, as shown in Table \ref{table-prediction-metrics-rank}. We found that C3 (Changed LOC of all files), C2 (Added LOC of all files), and C5 (The total number of lines of all files of the project currently) had high average rankings of 4.22, 4.78, and 4.78, respectively. This finding indicates that within the studied projects, metrics C3, C2, and C5 strongly influence the model's prediction performance. In contrast, C10 (Number of modified MPL files) and C11 (Whether to modify the source files of the main PL) had average rankings of 20.61 and 22.06, respectively, suggesting that metrics C10 and C11 have a relatively insignificant impact on the model's prediction performance in the studied projects.

\begin{center}
\fcolorbox{black}{gray!10}{\parbox{0.97\linewidth}{\textbf{Answer to RQ2:} Different metrics demonstrate varying predictive capabilities in the Random Forest predictions. According to the study results, C3 (Changed LOC of all files), C2 (Added LOC of all files) and C5 (The total number of lines of all files of the project currently) are the most crucial metrics in the JIT prediction of MPLBs.}}
\end{center}

\subsubsection{Performance of the JIT MPLB prediction models on reduced metric sets (RQ3)}

\begin{table*}[]
\caption{Model performance using selected metrics (RQ3)}
\scalebox{0.92}{
\centering
\begin{tabular}{@{\extracolsep{\fill}}cccccccccc}
\toprule
\textbf{Project} & \textbf{20\%AUC} & \textbf{ improved\%} & \textbf{40\%AUC} & \textbf{ improved\%} & \textbf{60\%AUC} & \textbf{improved\%} & \textbf{80\%AUC} & \textbf{ improved\%} & \textbf{100\%AUC} \\\midrule
        Airavata & 87.67 & 2.78 & 90.45 & 0.07 & 90.52 & -0.68 & 89.84 & 0.00 & 89.84 \\ 
        Ambari & 64.45 & 4.96 & 69.41 & 2.35 & 71.76 & 0.74 & 72.50 & 0.86 & 73.36 \\ 
        Atlas & 74.93 & 5.03 & 79.96 & -1.67 & 78.29 &2.22 & 80.51 & -0.54 & 79.97 \\ 
        Avro & 62.08 & 7.34 & 69.42 & 2.66 & 72.08 &-6.83 & 65.25 & 1.33 & 66.58 \\ 
        CloudStack & 72.16 & 4.06 & 76.22 & -0.88 & 75.34 &0.03& 75.37 & 0.50 & 75.87 \\ 
        Hawq & 74.83 & 1.67 & 76.50 & 0.83 & 77.33 &0.00& 77.33 & -7.08 & 70.25 \\ 
        Ignite & 69.67 & 1.50 & 71.17 & 0.45 & 71.62 &0.98& 72.64 & 0.42 & 73.06 \\ 
        Impala & 68.64 & 2.91 & 71.55 & -0.33 & 71.22 &1.06& 72.28 & 0.56 & 72.84 \\ 
        Kafka & 67.23 & 6.54 & 73.77 & -1.04 & 72.73 &1.25 & 73.98 & 1.05 & 75.03 \\ 
        Kudu & 70.48 & -0.30 & 70.18 & 3.82 & 74.00 &-0.68 & 73.32 & 1.48 & 74.80 \\ 
        Kylin & 69.38 & 6.09 & 75.47 & -3.13 & 72.34 &1.33 & 73.67 & 2.75 & 76.42 \\ 
        Mahout & 63.55 & 10.12 & 73.67 & -0.81 & 72.86 &1.72 & 74.58 & -0.66 & 73.92 \\ 
        Nifi & 71.46 & 6.30 & 77.76 & 1.44 & 79.20 &1.51& 80.71 & 1.77 & 82.48 \\ 
        Ranger & 75.32 & 2.36 & 77.68 & -1.08 & 76.60 &0.36& 76.96 & 0.09 & 77.05 \\ 
        Samza & 73.73 & 2.32 & 76.05 & 0.39 & 76.44 &0.52& 76.96 & -0.97 & 75.99 \\ 
        Storm & 77.36 & -4.50 & 72.86 & -1.54 & 71.32 &1.12& 72.44 & 3.50 & 75.94 \\ 
        Thrift & 68.78 & 2.68 & 71.46 & -0.77 & 70.69 &1.68 & 72.37 & 0.42 & 72.79 \\ 
        ZooKeeper & 73.44 & 6.30 & 79.74 & -0.14 & 79.60 &3.34& 82.94 & 1.27 & 84.21 \\ 
        Average & 71.40 & 3.78 & 75.18 & 0.04 & 75.22 &0.54 & 75.76 & 0.37 & 76.13                \\\bottomrule  
\end{tabular}
}
\label{table-selection-metrics}
\end{table*}

To answer RQ3, we sorted the 23 metrics based on the importance rankings obtained from RQ2. We started by selecting the top 20\% of the most important metrics for experimentation. Then, we gradually increased the number of metrics by a proportion of 20\% and conducted Random Forest predictions similar to RQ1. This was performed to further investigate whether the JIT MPLB prediction models could be simplified.

Table \ref{table-selection-metrics} presents the prediction performance generated using different percentages of metrics. We observed that (1) as more metrics were used, the average AUC gradually increased, and (2) after reaching 40\% of the metrics, the rate of AUC improvement became negligible, with the average AUC value across all projects only increasing by 0.04\% from 40\% to 60\% of the metrics. Therefore, we concluded that the model could be simplified using only 40\% of the metrics, and the simplified model could still perform effectively in the JIT MPLB prediction. 



\begin{center}
\fcolorbox{black}{gray!10}{\parbox{0.97\linewidth}{\textbf{Answer to RQ3:} When we used 40\% of the metrics, namely C3, C2, C5, C9, C6, C22, C19, C4, and C15, most of the models generated satisfactory prediction results (achieving an AUC of around 75\%). Increasing the number of metrics further did not significantly improve the model's prediction performance. }}
\end{center}

\subsubsection{Cross-project JIT MPLB prediction (RQ4)}
To answer RQ4, we performed two experiments using the dataset of the 18 projects. In the first experiment, we selected the data of one project from the dataset for model training and then used the trained model to predict the results for the remaining 17 projects. In the second experiment, since each project collected and utilized the same set of metrics, we used the data of 17 projects from the dataset of the 18 projects, trained a Random Forest prediction model, and finally tested the model using the one project that was not selected during training.

The results of the first experiment are shown in Table \ref{table-single-prediction}, where the number in a cell is the AUC of the project in the corresponding column as the test set for the trained model on the dataset of the project in the corresponding row. 
The last row of Table \ref{table-single-prediction} presents the average AUC of each project as the test set. Among the 18 projects, the average AUC of 13 projects is greater than 60.00\%, the average AUC of 5 projects falls in the range [50.00\%, 60.00\%). Through further calculations, we found that the mean value of the average AUC of each project as the test set is 62.05\%.

\begin{sidewaystable}[!htp]
\caption{AUC of cross-project JIT MPLB prediction models based on the dataset of a single project (RQ4)}
\centering
\scalebox{0.98}{
\resizebox{\textwidth}{!}{
\begin{tabular}{ccccccccccccccccccc}
\toprule
 \textbf{} & \textbf{Airavata } & \textbf{Ambari } & \textbf{Atlas } & \textbf{Avro } & \textbf{CloudStack } & \textbf{Hawq } & \textbf{Ignite } & \textbf{Impala } & \textbf{Kafka } & \textbf{Kudu } & \textbf{Kylin } & \textbf{Mahout } & \textbf{Nifi } & \textbf{Ranger } & \textbf{Samza } & \textbf{Storm } & \textbf{Thrift } & \textbf{ZooKeeper  } \\ \midrule
        Airavata & $\backslash$ & 52.13  & 49.87  & 52.59  & 60.79  & 53.41  & 58.70  & 49.95  & 50.81  & 51.13  & 53.29  & 49.87  & 55.92  & 50.29  & 50.13  & 57.45  & 50.21  & 50.72 \\ 
        Ambari & 81.36  & $\backslash$ & 65.53  & 71.47  & 67.15  & 64.62  & 70.55  & 65.64  & 66.18  & 54.36  & 65.84  & 61.93  & 71.09  & 67.61  & 65.21  & 65.79  & 66.40  & 72.64  \\ 
        Atlas & 53.66  & 53.35  & $\backslash$ & 80.22  & 53.36  & 67.94  & 59.08  & 62.83  & 65.83  & 66.50  & 55.49  & 57.19  & 63.31  & 70.49  & 73.76  & 65.06  & 68.17  & 74.72    \\ 
        Avro & 74.94  & 59.46  & 66.99  & $\backslash$ & 60.68  & 68.35  & 67.91  & 63.94  & 64.32  & 67.66  & 62.51  & 59.84  & 74.12  & 68.22  & 68.80  & 66.59  & 62.48  & 67.05    \\ 
        CloudStack & 85.46  & 62.37  & 57.20  & 58.05  & $\backslash$ & 72.94  & 67.90  & 61.25  & 60.69  & 58.96  & 66.89  & 57.94  & 71.69  & 61.94  & 54.63  & 58.79  & 59.96  & 60.36    \\ 
        Hawq & 47.00  & 63.60  & 67.97  & 69.41  & 64.60  & $\backslash$ & 67.51  & 68.14  & 66.57  & 63.66  & 66.51  & 62.81  & 63.08  & 67.08  & 63.09  & 61.80  & 66.83  & 63.06    \\ 
        Ignite & 79.47  & 64.30  & 66.37  & 74.54  & 63.85  & 73.86  & $\backslash$ & 68.60  & 66.84  & 64.86  & 66.49  & 61.95  & 73.82  & 67.93  & 65.50  & 62.99  & 68.23  & 70.93    \\
        Impala & 63.77  & 51.15  & 67.53  & 73.67  & 55.87  & 66.65  & 53.90  & $\backslash$ & 68.69  & 64.39  & 53.18  & 63.18  & 76.15  & 69.00  & 65.25  & 67.72  & 68.32  & 63.64    \\ 
        Kafka & 53.00  & 51.00  & 71.78  & 69.19  & 49.56  & 65.38  & 52.54  & 67.62  & $\backslash$ & 62.72  & 51.76  & 60.52  & 65.14  & 73.82  & 66.47  & 65.31  & 69.31  & 66.09    \\ 
        Kudu & 44.79  & 55.08  & 68.87  & 70.00  & 51.16  & 64.88  & 57.15  & 59.57  & 62.09  & $\backslash$ & 57.68  & 61.34  & 52.73  & 68.22  & 63.26  & 56.58  & 66.27  & 64.26    \\ 
        Kylin & 69.71  & 56.60  & 66.70  & 71.71  & 67.50  & 71.11  & 64.40  & 64.44  & 68.60  & 59.71  & $\backslash$ & 63.95  & 75.67  & 73.22  & 61.17  & 66.16  & 66.56  & 72.27    \\ 
        Mahout & 27.24  & 54.28  & 50.37  & 50.13  & 53.09  & 58.83  & 56.64  & 56.83  & 54.89  & 60.50  & 55.78  & $\backslash$ & 57.13  & 59.00  & 48.85  & 49.05  & 58.66  & 61.21    \\ 
        Nifi & 73.53  & 55.14  & 60.89  & 64.64  & 54.97  & 69.65  & 65.76  & 64.07  & 66.11  & 58.72  & 63.95  & 61.89  & $\backslash$ & 71.55  & 61.77  & 65.75  & 60.22  & 62.24    \\ 
        Ranger & 48.00  & 50.29  & 71.97  & 68.42  & 48.56  & 57.47  & 51.60  & 60.02  & 68.26  & 61.88  & 52.49  & 63.40  & 57.88  & $\backslash$ & 63.03  & 66.83  & 69.26  & 69.03    \\ 
        Samza & 48.88  & 49.72  & 66.06  & 77.54  & 49.94  & 52.42  & 50.04  & 52.78  & 57.25  & 63.13  & 50.88  & 50.21  & 50.43  & 55.88  & $\backslash$ & 54.07  & 59.16  & 68.69    \\ 
        Storm & 74.48  & 59.11  & 65.06  & 71.55  & 59.72  & 67.52  & 66.03  & 68.73  & 67.59  & 62.04  & 62.89  & 59.83  & 67.41  & 66.26  & 59.05  & $\backslash$ & 66.97  & 64.47   \\ 
        Thrift & 47.31  & 50.56  & 72.74  & 71.10  & 49.15  & 65.98  & 51.67  & 61.24  & 64.44  & 69.74  & 51.94  & 57.82  & 51.28  & 73.10  & 66.36  & 64.35  & $\backslash$ & 70.59    \\ 
        ZooKeeper & 49.45  & 49.90  & 65.60  & 78.64  & 49.00  & 52.55  & 50.00  & 52.51  & 54.05  & 59.78  & 50.42  & 51.76  & 50.22  & 54.56  & 67.09  & 62.22  & 61.50  & $\backslash$   \\ 
        Average & 60.12  & 55.18  & 64.79  & 68.99  & 56.41  & 64.33  & 59.49  & 61.66  & 63.13  & 61.75  & 58.12  & 59.14  & 63.36  & 65.77  & 62.55  & 62.15  & 64.03  & 66.00  \\\bottomrule         
\end{tabular}
}}
\label{table-single-prediction}
\end{sidewaystable}

In the second experiment, as shown in Table \ref{table-multi-prediction}, the AUC of most (14 out 18) projects (each as the test set) is larger than the average AUC of the corresponding projects in Table \ref{table-single-prediction}, and only 4 projects' AUC is less than the average AUC of the corresponding projects in Table \ref{table-single-prediction}. After calculation, we got that the average AUC of each project using the rest 17 projects as the training set is 66.18\%. This finding indicates that the AUC of the second experiment is significantly improved compared to the AUC of the first experiment.

However, certain projects, such as \textit{Ambari}, still did not perform well. We identified that this particular project had the highest proportion of bugs and that using it solely as a test set instead of a training set resulted in a substantial loss of bug-related information.

\begin{table}
\caption{AUC values of cross-project JIT MPLB prediction models based on the dataset of multiple projects (RQ4)}
\centering
\begin{tabular}{crcr}
\toprule
\textbf{Project} & \textbf{AUC\%} & \textbf{Project} & \textbf{AUC\%}  \\\midrule
Airavata         & 50.65             & Kudu             & 60.51             \\
Ambari           & 61.18             & Kylin           & 69.31             \\
Atlas            & 67.54             & Mahout            & 64.43             \\
Avro             & 68.55             & Nifi            & 79.51             \\
CloudStack       & 68.13             & Ranger           & 70.11             \\
Hawq             & 69.98             & Samza        & 65.57             \\
Ignite           & 71.15             & Storm             & 64.84             \\
Impala           & 62.91             & Thrift            & 66.57             \\
Kafka            & 67.33             & ZooKeeper           & 62.90
\\\bottomrule
\end{tabular}
\label{table-multi-prediction}
\end{table}


\begin{center}
\fcolorbox{black}{gray!10}{\parbox{0.97\linewidth}{\textbf{Answer to RQ4:} We found that using data from one project as the training set to predict MPLBs of other projects can yield decent results. When combining datasets from multiple projects, the prediction performance is significantly improved. However, the prediction performance of both approaches is still not as good as using data from a project to train the JIT MPLB prediction model for the project itself.}}
\end{center}

\section{Discussion}
\label{chap:discus}
In this section, we discuss JIT MPLB prediction performance in Section \ref{PredictionPerformance}, important metrics in JIT MPLB prediction in Section \ref{MPLSPL}, and prediction model parameter selection in Section \ref{ParameterSelection}.

\subsection{JIT MPLB Prediction Performance}
\label{PredictionPerformance}
To investigate the possible relationships between the project properties and the prediction performance, we calculated the Spearman correlation coefficient and \textit{p-value} for each property of the projects and the AUC values predicted by the Random Forest models in Table \ref{table-project-info}. The properties related to bug-inducing commits (i.e., the number and percentage of bug-inducing commits) were not included in the calculation, as the dataset had undergone the undersampling treatment and these two properties changed during the prediction process.
From Table \ref{think_relation}, we can observe that there is no significant correlation between these project properties and the AUC values, indicating that the given project properties are not associated with the JIT MPLB prediction effect.


\begin{table}
    \centering
    \caption{The correlation between the project properties and AUC.}
    \begin{tabular}{ccc}
    \toprule
        \textbf{} & \textbf{Correlation Coefficient} & \textbf{\textit{p-value}} \\ \midrule
        \#Commit & -0.1125 & 0.657 \\
        \#Ctbtr & -0.1641 & 0.515 \\
        Age & -0.0860 & 0.734 \\
        LOC & -0.0691 & 0.785 \\ 
        \#File & -0.0671 & 0.791 \\
        \#PL & -0.3692 & 0.132 \\
         \bottomrule
    \end{tabular}
    \label{think_relation}
\end{table}


In the results of RQ3 (see Table \ref{table-selection-metrics}), we noticed that the AUC values of projects \textit{Hawq} and \textit{Storm} decrease when all prediction metrics were used compared with using only 20\% of the prediction metrics. This implies that not all prediction metrics are beneficial to the prediction performance. Specifically, for \textit{Hawq}, the AUC value increases when the percentage of prediction metrics used increases from 20\% to 80\%, but decreases significantly when using all the prediction metrics; for \textit{Storm}, the AUC value decreases when the percentage of prediction metrics used increases from 20\% to 60\%, and increases when the percentage of prediction metrics used increases from 60\% to 100\%. Therefore, it is necessary to select different prediction metrics for different projects.


\subsection{Important Metrics in JIT MPLB Prediction}\label{MPLSPL}
In the results of RQ2, three metrics, i.e., C3 (Changed LOC of all
files), C2 (Added LOC of all files), and C5 (The total number of lines of all files of the project currently), are the most crucial metrics in the JIT prediction of MPLBs, have the most important impact on the prediction results. We further observed that the importance ranking scores of C9 and C22 are 4.89 and 5.83 respectively, which are quite close to the scores of the first three metrics, hence C9 and C22 also play an important role in JIT MPLB prediction.

However, when selecting C1-C7 (language-independent metrics in Category 1) only as the dataset and using Random Forest for training and prediction, we obtained an average AUC value of 73.07\% for all projects (as shown in Table \ref{table-spl-metric-prediction}), which is merely 3.06\% smaller than the average AUC value of 76.13\% when using all the 23 prediction metrics. It seems that the metrics specific to MPL projects (i.e., C8-C23) do not play as important roles as demonstrated by their importance ranking scores in Table \ref{table-prediction-metrics-rank}, where metrics such as C9, C22, and C19 have relatively high rankings. One possible reason is that some metrics specific to MPL projects are close to corresponding language-independent metrics. For instance, C9 (Added LOC of MPL files) is close to C2 (Added LOC of all files) in most cases in a project, which is confirmed by their similar importance ranking scores as shown in Table \ref{table-prediction-metrics-rank}.

In traditional non-JIT bug prediction, certain metrics, such as LOC (equivalent to C5 in our study), significantly influence prediction results. Among these, LOC stands out as a cost-effective and readily obtainable metric, proving to be one of the most effective predictors of bugs across various scenarios~\citep{menzies2002metrics}. This implies that the project size in LOC plays a critical role in both JIT MPLB and non-JIT bug prediction.



\begin{table}
\caption{AUC value of the JIT MPLB prediction based on the dataset containing only the metrics in Category 1}
\centering
\begin{tabular}{crcr}
\toprule
\textbf{Project} & \textbf{AUC\%} & \textbf{Project} & \textbf{AUC\%}  \\\midrule
Airavata         & 86.30             & Kudu             & 70.95             \\
Ambari           & 67.40             & Kylin           & 70.15             \\
Atlas            & 78.32             & Mahout            & 66.70             \\
Avro             & 68.75             & Nifi            & 75.38             \\
CloudStack       & 74.91             & Ranger           & 75.05             \\
Hawq             & 73.58             & Samza        & 74.55             \\
Ignite           & 71.44             & Storm             & 75.27             \\
Impala           & 69.79             & Thrift            & 71.90             \\
Kafka            & 70.54             & ZooKeeper           & 74.34 \\ \hline 
\textbf{Average} & \textbf{73.07} & &
\\\bottomrule
\end{tabular}
\label{table-spl-metric-prediction}
\end{table}

\subsection{Prediction Model Parameter Selection}
\label{ParameterSelection}
In all the experiments in this study, the parameter for the random seed has been set to the same value, i.e., 42. The purpose is to facilitate better reproducibility of the study for future researchers. Other functions that require the \textit{random\_state} parameter are also set to 42 for the same purpose. In the context of the Random Forest experiments, the number of trees in the forest was set to 100. In this work, the default value of 100 for \textit{n\_estimators} in the scikit-learn library's \footnote{https://scikit-learn.org/stable/index.html} Random Forest models is chosen to achieve a balance between prediction performance and efficiency. The increase in the number of trees has a positive impact on the performance of the prediction results. Researchers using Random Forest prediction models can consider adjusting the number of trees to make a tradeoff between required performance and operational efficiency.

\section{Threats to Validity}\label{chap:threats}

We discuss the potential threats to the validity of this study according to the guidelines proposed in~\citep{runeson2009guidelines}.

\subsection{Threats to Construct Validity} 
First, we cannot guarantee that all metrics we proposed can comprehensively represent all information in the code. However, the selection of 23 prediction metrics is based on previous research in the field~\citep{hassan2009predicting, moser2008comparative, nagappan2005use, menzies2002metrics} and our further exploration, with the aim of covering as many metrics as possible that might lead to bugs. The prediction results also demonstrate the effectiveness of these metrics in JIT MPLB prediction. Furthermore, our approach is highly extensible in the sense that incorporating new metrics into our models would be straightforward. 

Second, a potential threat may come from the data collection process. Any inaccuracies in the collected metrics could affect the performance of the prediction results. 
We obtain raw data from JIRA and GitHub, utilizing the Pydriller tool to extract required metrics. These are commonly used data sources and data collection tools among researchers~\citep{abidi2021multi, tsoukalas2022td, inoue2022impact}. Besides, we have provided a detailed data collection process (see Section \ref{Collection-steps}) and code~\citep{dataset} to ensure the standardization of our data collection procedure. 
Through these steps, we tried to ensure that all metrics were collected with minimized biases.

Finally, there are some PLs (e.g., Perl) that are used in the 18 selected Apache MPL OSS projects but not supported by the Lizard tool, which results in the loss of data regarding such PLs when calculating the second category of metrics. However, considering that 14 and 3 projects have more than 91.4\% and 83.8\% respectively of the source code covered by the PLs supported by Lizard, and only one project has less than 83.8\% (i.e., 77.6\%) of its source code covered by the PLs supported by Lizard, this threat is limited.

\subsection{Threats to Internal Validity} 
A threat arises from the coverage of the bugs in the selected MPL software systems. To be specific, not every bug-fixing commit has included the key of the bug in the commit message, and thus some MPLBs are not included in our dataset. We consider all commits within the specified time period to mitigate this threat.

\subsection{Threats to External Validity} 
In this work, we used 18 MPL OSS projects from the Apache OSS ecosystem, and we cannot claim that the research results are generalizable for MPL projects from other OSS ecosystems. 
To mitigate this threat, we selected systems that are continuously evolving and have long histories. These systems have received updates over the past year and vary in popularity, size, category, and number of commits. Furthermore, the dataset we considered and collected was comprehensive, covering all commits from the inception of each selected MPL software system to the cut-off date of data collection, providing fine-grained data.

\subsection{Threats to Reliability} 
A potential threat lies in the tools used for calculating prediction model metrics and evaluating experimental results. These tools were mainly implemented by the second author, while the code and key functions were reviewed regularly by the first author. We have provided the relevant code~\citep{dataset} for other researchers to reproduce the experiments conducted in this study. Another threat is related to the dataset used in this study. To improve the reliability of the research results, we also provide the dataset used, as well as the intermediate results~\citep{dataset}. Thus, with these measures, the threats to reliability are reduced.

\section{Conclusions and Future Work}
\label{conclusions}
In this work, we introduced a set of prediction metrics, utilized several machine learning algorithms to construct prediction models for the JIT prediction of MPLBs, and found that Random Forest performed the best. Subsequently, we analyzed the importance of the metrics and ranked these metrics. To reduce the complexity of JIT MPLB prediction models, we selected the most crucial metrics to build simplified JIT MPLB prediction models. Lastly, we evaluated the performance of the cross-project JIT MPLB prediction models.

From the evaluation of our JIT MPLB prediction with 18 Apache MPL OSS projects, we have demonstrated: (1) we can construct robust JIT MPLB prediction models based on our selected metrics and the Random Forest algorithm, achieving an average AUC of around 76\%; (2) we can build simplified JIT MPLB prediction models using the top 40\% ranked metrics; (3) cross-project JIT MPLB prediction is feasible, and the size of the training dataset has a positive impact on prediction performance.

This research on JIT MPLB prediction helps raise the practitioners' awareness of MPLB prediction and prevention given the higher bug fixing cost than SPL bugs in general~\citep{li2023understanding2, li2023understanding}. This study also provides practitioners with an effective  means to get informed timely whether a code change introduces MPLBs. For researchers, this study has confirmed the feasibility of within-project and cross-project JIT MPLB prediction, which has opened a new research branch of defect prediction. In addition, the results of this study provides researchers with a baseline for JIT MPLB prediction research. To our knowledge, our work is the first exploration in MPLB prediction, and there is still a lot of room for further development in the future to optimize prediction metrics as well as prediction algorithms, and finally improve the performance of MPLB prediction.

There are several aspects that can be further explored in the next step. 
(1) We plan to collect more potentially relevant metrics to improve the AUC of JIT MPLB prediction. Specifically, we will further explore potential metrics at the function and class levels that may have an impact on the prediction. 
(2) We are also interested in predicting MPLBs in the MPL software systems written in a specific combination of programming languages, such as Java and C++. This allows us to collect language-specific features to predict MPLBs and may achieve better prediction performance. 
(3) We will adapt our JIT MPLB prediction approach in closed-source commercial MPL software projects to enhance the practicality of the approach in a wider scope. To be specific, we plan to collaborate with industrial partners, collect the needed prediction metrics and label commits in their projects, and replicate the prediction experiments in an industrial setting. 

\section*{Data availability}
We have shared the link to our dataset in the reference~\citep{dataset}.

\section*{Acknowledgments}
This work was funded by the National Natural Science Foundation of China under Grant Nos. 62176099 and 62172311, the Natural Science Foundation of Hubei Province of China under Grant No. 2021CFB577, and the Knowledge Innovation Program of Wuhan-Shuguang Project under Grant No. 2022010801020280.

\printcredits

\appendix
\section*{Appendix}
A brief description for each selected MPL OSS Apache project is shown in Table \ref{table-project-description}, 
\begin{table*}
    \centering
    \caption{Project description}
    \begin{tabular}{cc}
    \toprule
        \textbf{Project} & \textbf{Description} \\ \midrule
        Airavata & An open-source distributed computing
and data management platform \\ 
        Ambari & An open-source cluster management and deployment tool \\ 
        Atlas & An open-source data governance and metadata management platform \\ 
        Avro & A data serialization framework \\ 
        CloudStack & An open-source cloud infrastructure management platform \\ 
        Hawq & An open-source SQL query engine \\ 
        Ignite & An open-source in-memory computing platform \\ 
        Impala & An open-source distributed SQL query engine \\ 
        Kafka & A high-performance, distributed streaming data platform \\ 
        Kudu & An open-source distributed storage engine \\ 
        Kylin & An open-source big data analytics engine \\ 
        Mahout  & An open-source machine learning library \\ 
        NiFi & An open-source data integration tool \\ 
        Ranger & An open-source access control and security management framework \\ 
        Samza & An open-source stream processing framework designed to process real-time data streams \\
        Storm & An open-source real-time computing system for distributed streaming data processing \\ 
        Thrift  & A cross language remote procedure call framework and data serialization framework \\ 
        ZooKeeper & A distributed coordination service \\ \bottomrule
    \end{tabular}
    \label{table-project-description}
\end{table*}

\bibliographystyle{cas-model2-names}

\bibliography{references}

\balance

\end{sloppypar}
\end{document}